\documentclass[a4paper]{spie}  %>>> use this instead for A4 paper
\usepackage[a4paper, left=1.93cm, right=1.93cm, top=2.54cm, bottom=4.94cm]{geometry}
\usepackage{amsmath,amsfonts,amssymb}
\usepackage{graphicx}
\usepackage[colorlinks=true, allcolors=blue]{hyperref}
\usepackage{booktabs}
\usepackage{caption}
\usepackage{cancel}

\title{Low-cost, ultra-wideband, differential low-noise amplifier for interferometric radio telescopes}

\author[a,b]{Sophia Da Costa}
\author[b]{Albert Wai Kit Lau}
\author[a,b]{Keith Vanderlinde}
\affil[a]{David A. Dunlap Department of Astronomy and Astrophysics, University of Toronto, 50 St George St, Toronto, Canada}
\affil[b]{Dunlap Institute of Astronomy and Astrophysics, University of Toronto, 50 St George St, Toronto, Canada}

\authorinfo{Send correspondence to S. Da Costa\\E-mail: sophia.dacosta@mail.utoronto.ca}

\begin{document} 
\maketitle

\begin{abstract}

We present a low-cost, ultra-wideband, dual-polarized differential low-noise amplifier that eliminates the need for a pre-amplification balun. Built entirely from commercial surface-mount components, the design costs approximately $\$20$ USD per unit. The amplifier has been characterized through laboratory gain and noise measurements together with feed-coupled hot-cold testing at the Dominion Radio Astrophysical Observatory. A noise figure close to 0.3 dB is demonstrated across a 10:1 bandwidth when matched to a constant 130~$\Omega$ source, while feed-coupled measurements yield system noise temperatures as low as 25 K when matched to a Vivaldi feed operating over a 300--1500 MHz bandwidth.
\end{abstract}


\section{INTRODUCTION} %%%%%%%%%%%%%%%%%%%%%
\label{sec:intro} 

Next generation radio telescopes will be making use of large-N, small-D architectures to conduct sensitive, wide field surveys. Up-and-coming observatories such as DSA, CHORD and the SKA serve as examples that will be using ultra-wideband (UWB) technology to conduct observations at broad bandwidths \cite{hallinan2019,vanderlinde2019, dewdney2009}. This will enable a large parameter space for 21cm intensity mapping and transient searches such as the discovery and localization of fast radio bursts. Projects such as these require the design of robust signal chains that are both low-cost and low-noise while maintaining ultra-wideband capabilities. 

Many feeds used in radio astronomy are inherently differential, naturally exhibiting polarity, in which the received signal is represented by two equal and opposite voltages carried on independent conductors. This contrasts with single-ended electronics, in which the signal is represented by the voltage on one conductor relative to ground and is often designed around a 50~$\Omega$ impedance standard. Consequently, differential feeds are commonly interfaced with single-ended electronics using a balun. However, the addition of a balun can increase the system noise temperature, limit the achievable bandwidth, and introduce cross-talk between dual-polarized channels \cite{cappallo2025, bhaumik2009}. Together with the signal attenuation and cost associated with conventional 50~$\Omega$ coaxial cables, these limitations have motivated the development of differential signal transmission techniques \cite{lau2024}. By employing a differential low-noise amplifier (LNA), the feed can be connected directly to the receiver chain without an intermediate balun, making differential architectures an attractive alternative to conventional single-ended systems.

Several differential LNA architectures have been proposed for radio astronomy applications \cite{perez2011,jimenez2019,tan2000, sutinjo2015, ellingson2013}. While these designs demonstrated the advantages of directly coupling differential LNAs to balanced feed elements, many rely on custom amplifier topologies, specialized fabrication techniques, or lack the bandwidth required for emerging radio astronomy instruments. Commercial differential amplifiers are also available; but few provide the combination of low noise figure and ultra-wideband performance for sensitive radio astronomy applications. When suitable devices are available, they are often costly relative to large-N telescope deployments. In contrast, recent advances in the telecommunications industry have produced a number of single-ended amplifiers with excellent wideband noise and gain performance. One example is the Qorvo QPL9547, an ultra-low-noise amplifier housed in a 2.0 mm $\times$ 2.0 mm surface-mount package and available for approximately $\$$2 USD. While the QPL9547 is inherently a single-ended device, two amplifiers can be configured as a differential pair, allowing each one to independently amplify one polarity of the differential input signal.

Motivated by these considerations, we present a low-cost, dual-polarized differential LNA with UWB capabilities. The design costs approximately $\$$20 USD per LNA, and the exclusive use of commercially available surface-mount components facilitates mass production. The amplifier employs a multi-stage architecture with a Qorvo QPL9547 serving as the first stage, resulting in a compact design measuring 5 cm $\times$ 5 cm $\times$ 3 cm (L $\times$ W $\times$ H). The LNA achieves a minimum noise temperature of 17 K at 900 MHz when presented with an optimum differential source impedance of 135~$\Omega$. To demonstrate the application of the design, the LNA was integrated with a tapered-slot Vivaldi feed with a differential input impedance of approximately 110~$\Omega$ over a 300--1500 MHz bandwidth \cite{mackay2023}.

The remainder of this paper is organized as follows. Section~\ref{sec:method} describes the LNA topology, physical implementation, and the measurement procedures used to characterize its gain and noise performance. Section~\ref{sec:results} presents the measured gain, temperature stability, and noise temperature of the LNA. Section~\ref{sec:discuss} summarizes the key performance drivers and discusses future developments and applications of the design.

\section{METHODS} %%%%%%%%%%%%%%%%%%%
\label{sec:method}
\subsection{Design Considerations}

The primary objective of the LNA design was to provide a low-cost differential front-end suitable for integration with modern ultra-wideband radio astronomy feeds. Initial development was carried out using the tapered-slot Vivaldi feed presented in Ref.~\citenum{mackay2023}, which exhibits a differential input impedance of approximately 110~$\Omega$ over a 300--1500 MHz bandwidth. Generally, the LNA was designed to provide a suitable differential noise match given an input impedance close to $\sim100~\Omega$ across a 10:1 bandwidth while enabling direct connection to the feed without the use of an intermediate balun.
In addition to noise performance, affordability and scalability were key design requirements. Future radio astronomy instruments increasingly employ large-N architectures consisting of hundreds to thousands of receiving elements, making the cost and manufacturability of front-end electronics an important consideration. To address this, the design was implemented exclusively using commercially available surface-mount components and standard low-cost printed circuit board (PCB) substrates which minimizes production cost to support large-scale deployments. Another design objective was to maintain compatibility with conventional 50~$\Omega$ measurement and receiver systems. To achieve this, the LNA was implemented using a differential-input, single-ended-output (DISO) architecture, which is widely used in applications requiring direct integration of differential feed elements with standard single-ended RF infrastructure \cite{sutinjo2015, tan2000, ellingson2013, fan2022}. However, if a differential receiver chain is available downstream, the output balun can be removed to allow the LNA to operate with a fully differential output.

\subsection{Differential Topology}

\subsubsection{Architecture}

\begin{figure}[hbt!]
    \centering
    \includegraphics[width=0.7\linewidth]{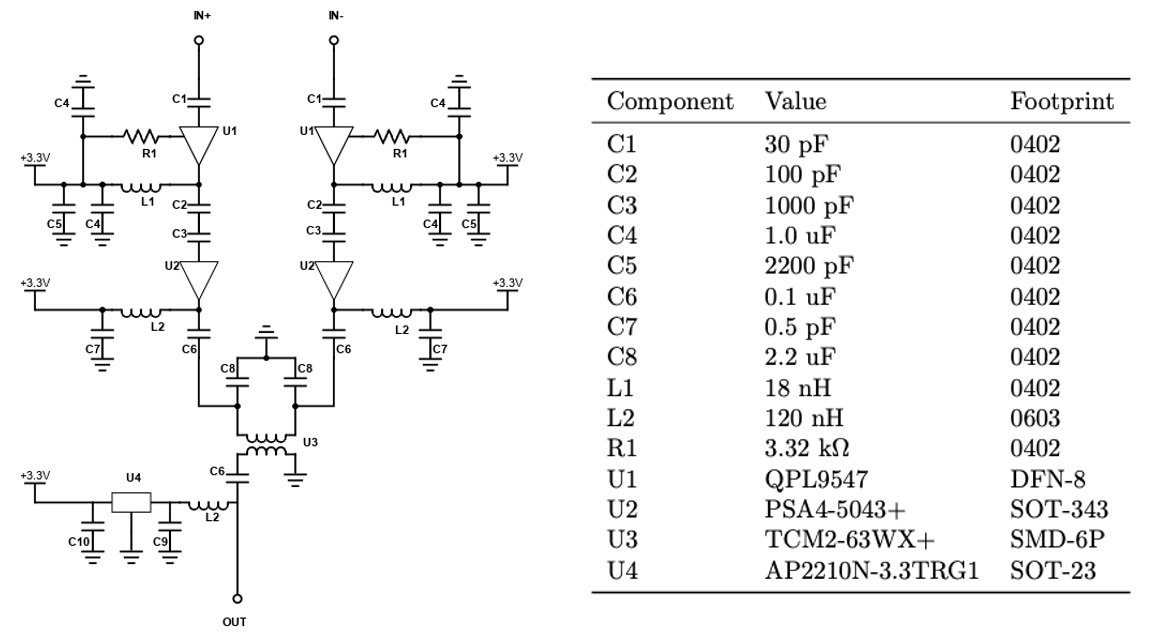}
    \caption{Electrical schematic of a single polarization of the differential LNA prototype. Each polarization employs a differential front-end architecture followed by a balanced-to-unbalanced conversion stage, resulting in a 50~$\Omega$ single-ended output. Component values and package footprints are listed in the accompanying table. }
    \label{fig:circut}
\end{figure}

% topolgy + circut 
The LNA prototype employs commercially available single-ended surface-mount components arranged in a differential topology consisting of two identical amplifier branches. Each branch independently amplifies one polarity of the differential input signal with respect to a common ground. The amplifier uses two gain stages, after which the outputs are combined using a balun to provide compatibility with conventional 50~$\Omega$ receiver chains. As a result, each polarization channel of the LNA presents a differential input and a single-ended output. The bandwidth of the prototype is primarily limited by the output balun which is specified for operation up to 4 GHz. The active amplifier stages support operation to approximately 6 GHz, meaning that higher-frequency operation could be achieved by using a different broadband conversion stage or by removing the balun entirely to maintain a fully differential circuit. The schematic for a single polarization channel is shown in Figure \ref{fig:circut}, with the corresponding component values listed. The design operates from a 3.3~V supply and draws approximately 150~mA per polarization resulting in a total power consumption of approximately 1~W for the dual-polarized LNA.

\begin{figure}[hbt!]
    \centering
    \includegraphics[width=0.65\linewidth]{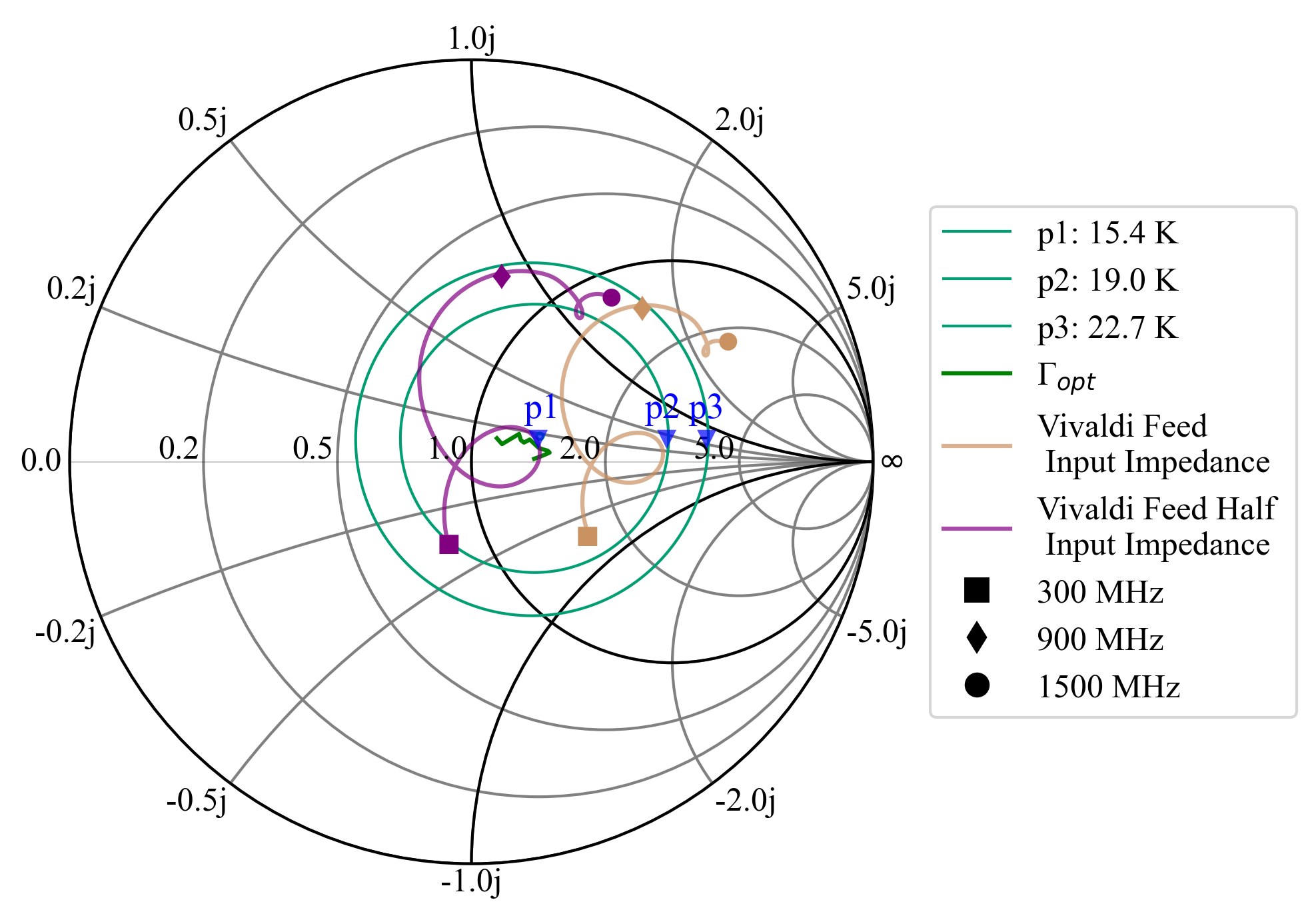}
    \caption{Smith chart illustrating the simulated differential input impedance of the Vivaldi feed and the noise matching characteristics of the QPL9547 LNA. (Turquoise) Noise circles of QPL9547 LNA at 900 MHz, with the central point of the innermost circle indicating the optimal noise match for the LNA and each subsequent circle indicating an increase in noise temperature. (Green) $\Gamma_{opt}$ of QPL9547 across 300--1500 MHz band, this remains relatively stable across ultra-wide bandwidths. (Brown) Simulated differential input impedance of Vivaldi feed within 300--1500 MHz frequency range. (Purple) Simulated differential input impedance of Vivaldi feed halved, representing the reference match to one amplifier branch of the LNA. (This data is presented on a Smith chart normalized to 50~$\Omega$).}
    \label{fig:noise_circ}
\end{figure}

In this configuration, each amplifier branch is modeled as seeing half of the differential source impedance. Therefore, the noise performance of the differential LNA can be evaluated by considering a single amplifier branch matched to half of the desired differential input impedance. Figure \ref{fig:noise_circ} compares the optimum noise match of the QPL9547 with half of the differential input impedance of the Vivaldi feed reported in Ref.~\citenum{mackay2023}. At 900 MHz, the QPL9547 exhibits an optimum noise match with an input impedance of 67~$\Omega$, corresponding to a differential optimum noise match of 135~$\Omega$ when two amplifiers are configured in the differential topology. Simulations across the QPL9547 operating bandwidth of 0.1--6 GHz indicate that $\Gamma_{opt}$ remains relatively stable, ranging from approximately 100~$\Omega$ to 140~$\Omega$. As the QPL9547 serves as the first-stage amplifier these results suggest that the proposed design is well matched to differential feeds with input impedances close to 100~$\Omega$.

\begin{figure}[hbt!]
    \centering
    \includegraphics[width=0.9\linewidth]{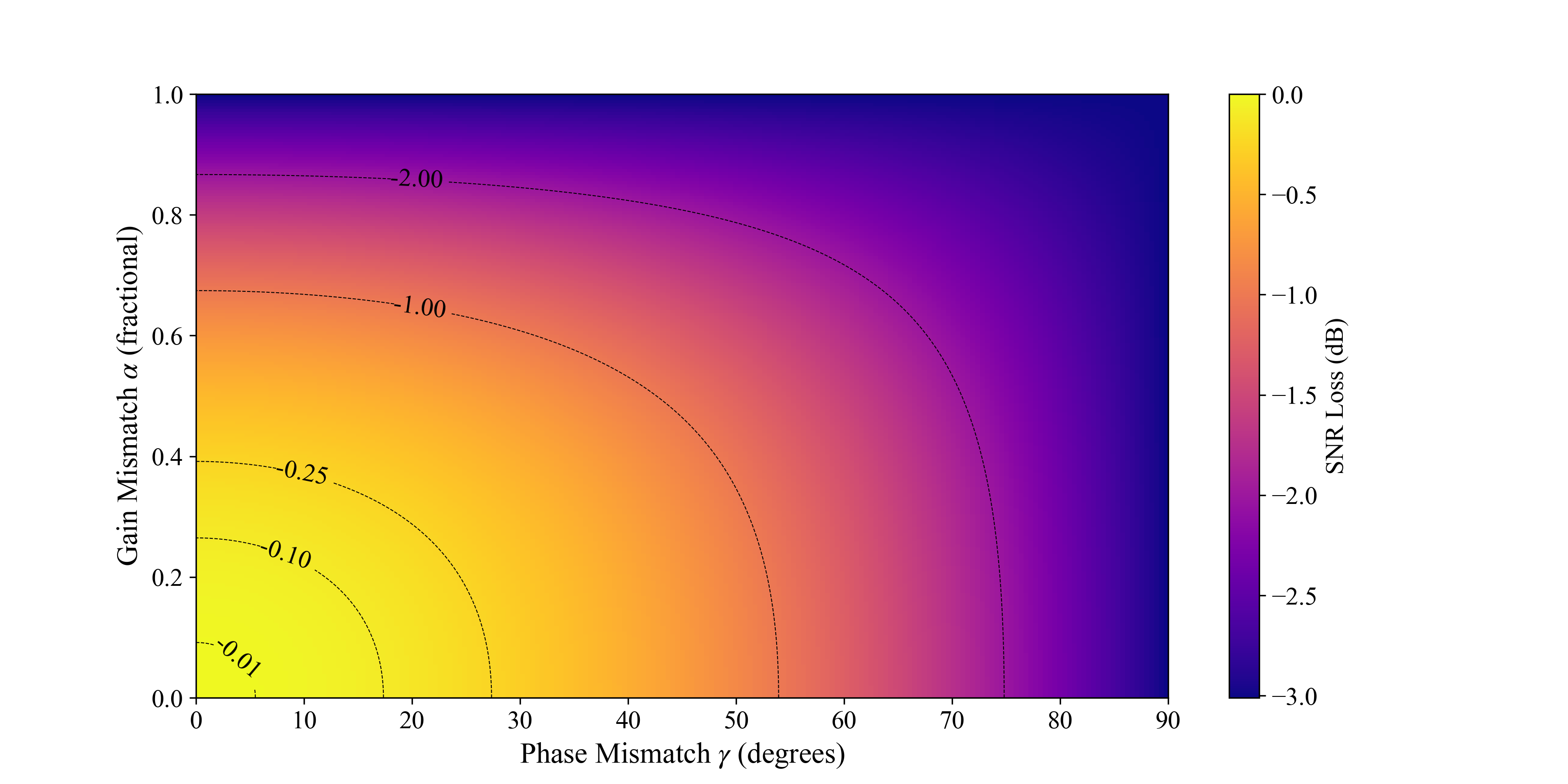}
    \caption{SNR degradation of a differential system relative to an ideal single-ended system as a function of gain and phase mismatch between the two amplifier branches (polarities). The contours indicate the expected reduction in SNR due to component tolerances.}
    \label{fig:snr_deg}
\end{figure}

Employing two amplifiers side-by-side in this configuration exhibits the same input-referenced noise performance as an equivalent single-ended system provided the two branches have identical gain, phase, and noise characteristics. In this case the differential signal adds coherently while the amplifier noise contributions remain uncorrelated. This results in an equivalent system noise that is unchanged relative to the single-ended circuit. A derivation of this relationship is presented in Appendix~\ref{sec:noise_derivation}. In practice, gain and phase mismatches arise between amplifier branches due to component tolerances and manufacturing variations. To quantify these effects the signal-to-noise ratio (SNR) of the differential architecture was derived as a function of gain and phase imbalance. The resulting SNR degradation relative to an ideal single-ended system is summarized in Figure~\ref{fig:snr_deg}, demonstrating the tolerance of the topology to realistic amplifier mismatches. The full derivation is provided in Appendix~\ref{sec:snr_derivation}.

\subsubsection{PCB implementation}

\begin{figure}
    \centering
    \includegraphics[width=0.9\linewidth]{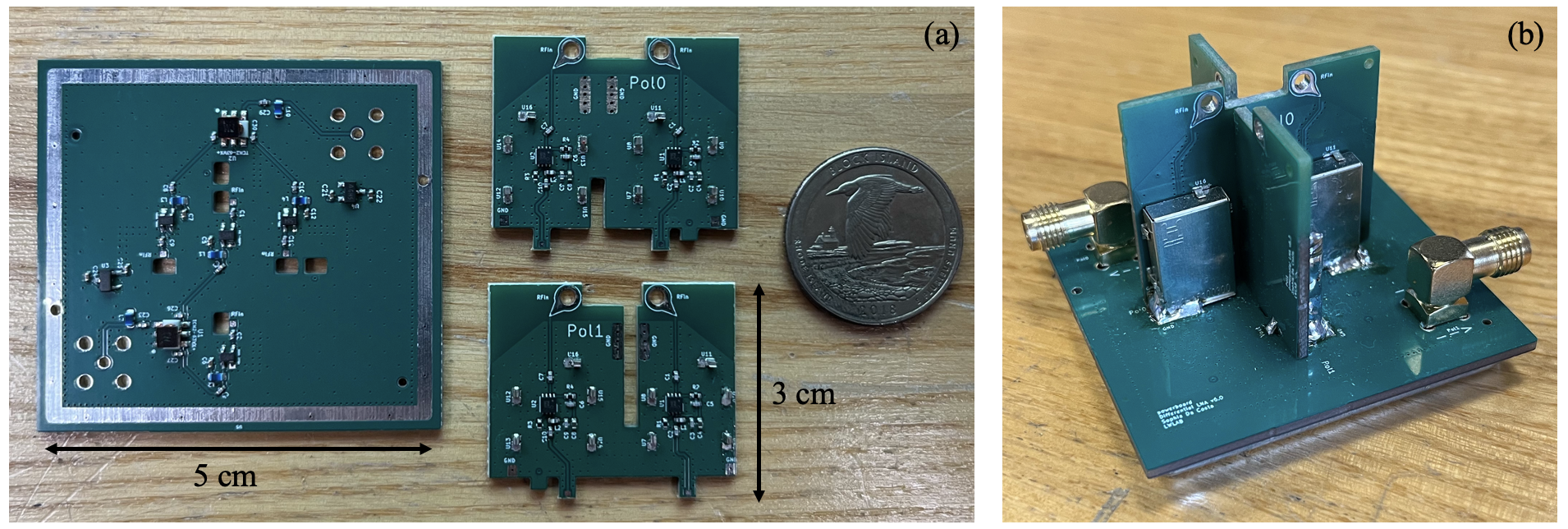}
    \caption{(a) Unassembled and (b) assembled dual-polarized differential LNA prototype. The assembly consists of three PCBs fabricated on standard FR-4 substrate. These include two input boards (one for each polarization) measuring (3 $\times$ 3) cm and housing the first-stage QPL9547 amplifiers. The input boards slot together and connect directly to the larger power board which houses the second-stage amplifiers, voltage regulators, and output baluns. Individual RF shields are installed over each polarization on the input boards, while a larger RF shield encloses the power board.}
    \label{fig:layout_PCB}
\end{figure}

The LNA is implemented using three PCBs fabricated on a standard FR-4 substrate with a thickness of 1.6 mm and dielectric constant of $\epsilon_r = 4.4$. The transmission-line layout employs a grounded coplanar waveguide (GCPW) architecture which provides high isolation and reduced radiative losses \cite{hu2024}. Figure \ref{fig:layout_PCB} shows both the assembled and unassembled dual-polarized model.
The first amplification stage is implemented on two differential input boards, one for each polarization, which physically slot together. Each input board contains a QPL9547 first-stage amplifier located close to the feed terminals to minimize losses. These boards connect to a larger power board that houses the second-stage amplifiers, voltage regulators, and baluns. To reduce electromagnetic coupling and external interference, RF shields are installed over each polarity on the two input boards and a larger shield is installed on the power board. The complete assembly measures approximately 5 cm $\times$ 5 cm $\times$ 3 cm (L $\times$ W $\times$ H) with an estimated fabrication cost of approximately $\$$20 USD.

The interface between the three PCBs required careful grounding to maintain consistent RF performance. In particular, all components associated with a given polarization are required to share a common ground reference despite the presence of slots between the interconnected boards. To preserve a low-impedance AC return path additional ground connections were introduced across these discontinuities to ensure that return currents followed the shortest possible path. Without these connections return currents were forced to propagate around the slots which increased the effective path length and produced impedance discontinuities. During initial testing these effects manifested as reflections in the measured S$_{21}$.

\subsubsection{Feed interface and mounting}
\begin{figure}[hbt!]
    \centering
    \includegraphics[width=0.9\linewidth]{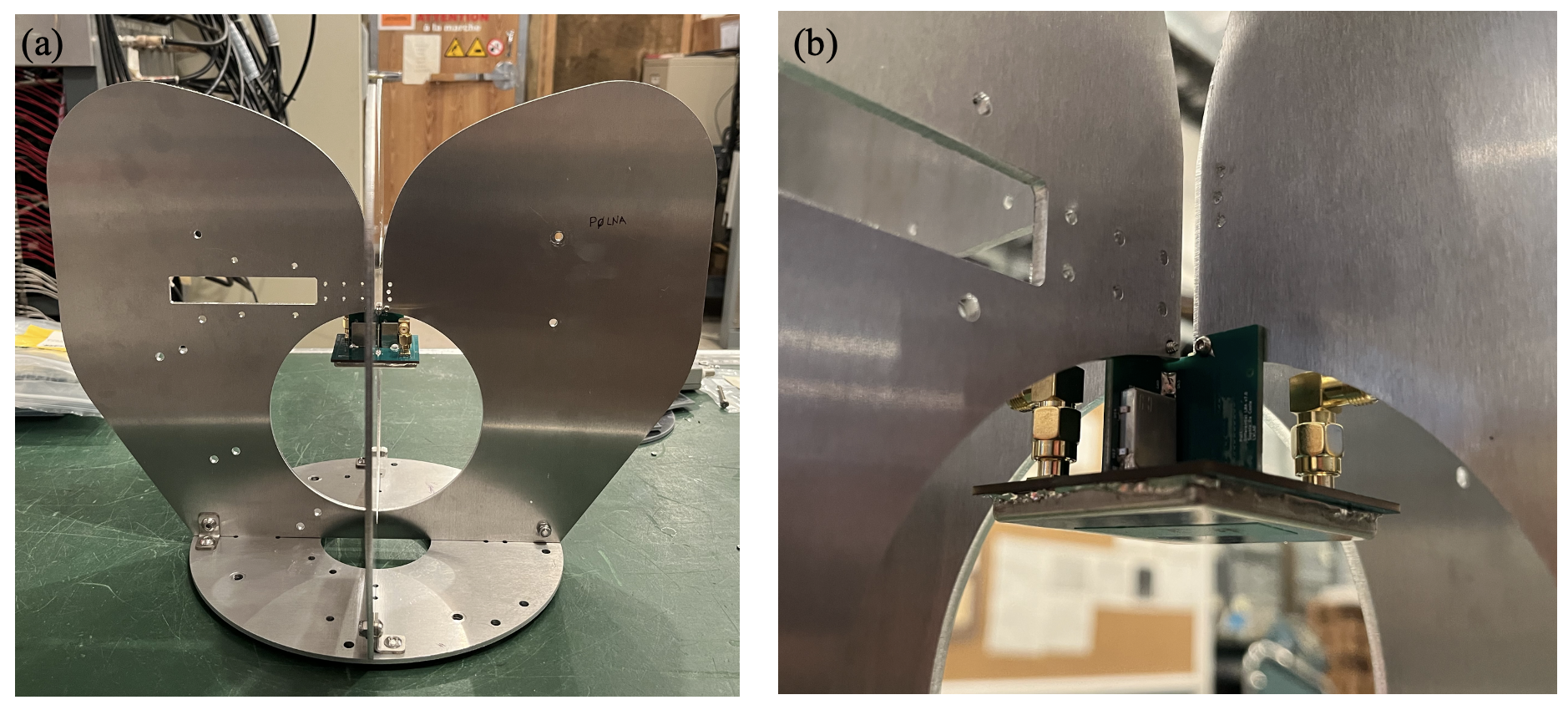}
    \caption{(a) Differential LNA mounted directly on the dual-polarized Vivaldi feed. (b) Close-up of the feed–LNA interface showing the M2 connection points used to attach the LNA to the feed. The mounting arrangement places the LNA within the feed backshort (large circular cavity), reducing electromagnetic coupling between the feed and the amplifier. }
    \label{fig:feed_mount}
\end{figure}

% mounting structure and teardrop and cuts to PCB 
The LNA is mounted directly to the Vivaldi feed using 2 mm screws located at the four differential input terminals, which make electrical contact through teardrop-shaped connection pads. Electromagnetic simulations indicated that teardrop shaped pads induced minimal shifts in input impedance due to their ability to act as an impedance taper, gradually changing in width, minimizing impedance discontinuities from feed to waveguide. The ground plane is also cut back along the upper edges of the PCB to reduce electromagnetic coupling between the feed and LNA. With this mounting arrangement the lower portion of the LNA does not make direct contact with the feed structure and instead resides within the backshort cavity. This allows the amplifier to be integrated directly at the feed terminals while maintaining both a compact footprint and minimizing interference with the active regions of the feed which can be seen in Figure~\ref{fig:feed_mount}.

\subsection{Measurements}

\subsubsection{50-$\Omega$ measurements}

\begin{figure}[hbt!]
    \centering
    \includegraphics[width=0.9\linewidth]{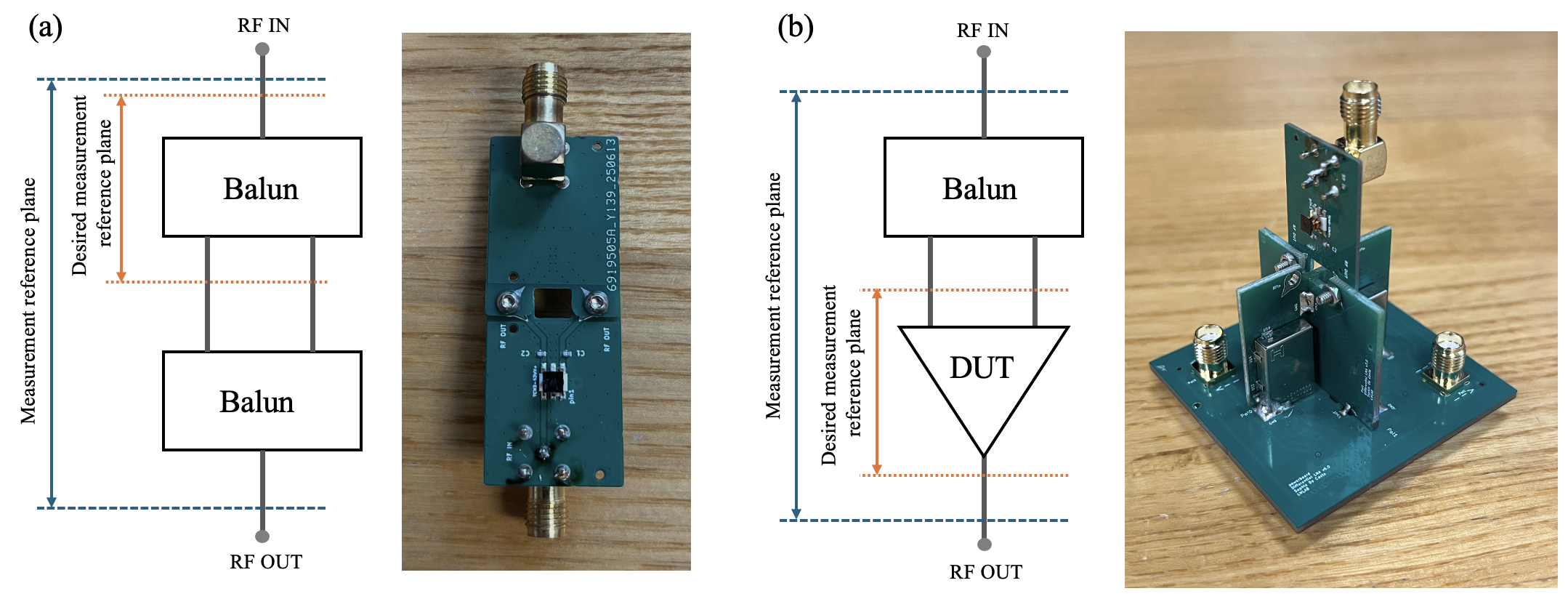}
    \caption{50~$\Omega$ measurement configurations used for (a) characterizing the insertion loss of the input balun using two identical balun boards connected back-to-back, and (b) measuring the gain ($S_{21}$) of the differential LNA.}
    \label{fig:gain_setup}
\end{figure}

The characterization of differential devices using conventional laboratory equipment is non-trivial as it is commonly designed for single-ended systems. To facilitate gain measurements the differential LNA was integrated into a fully 50~$\Omega$ test configuration. This was done by incorporating a balun at the input of the amplifier (see Figure~\ref{fig:gain_setup}), similar to the approach adopted in Ref.~\citenum{perez2011}. The LNA presented in this work incorporates an output balun as part of the circuit architecture, requiring only a single external balun at the input. To provide a suitable interface a dedicated balun board was developed using a Mini-Circuits TCM2-63WX+ transformer implemented on an FR-4 substrate. The board uses the same connection ports as the LNA allowing the direct attachment to the LNA input terminals using M2 screws.
The gain ($S_{21}$) was measured using a vector network analyzer. To isolate the response of the LNA the insertion loss of the input balun was independently characterized using two identical balun boards connected back-to-back as shown in  Figure~\ref{fig:gain_setup}. Half of the measured insertion loss was then removed from the $S_{21}$ response.

The same measurement setup was used to evaluate gain stability under varying environmental conditions. Temperature testing was conducted using an in-house programmable environmental chamber with a 1 m-sided interior. A dedicated cable feedthrough allowed RF and power connections to be routed to the device under test while maintaining temperature control. The chamber supports operating temperatures from $-65^{\circ}$C to $+150^{\circ}$C and enables automated temperature cycling. Measurements were conducted between $-40^{\circ}$C and $+60^{\circ}$C to quantify gain variations across the LNA operational band under conditions representative of those expected during deployment. The same test was carried out with the input balun isolated to ensure its contribution could be removed from the final measurement. 

\subsubsection{Noise measurements}
The noise temperature of the LNA was characterized using the Y-factor method, a standard technique for determining the noise contribution of RF devices. The method compares the output power of the system when connected to hot and cold sources of known temperature. The ratio of the measured powers, referred to as the Y-factor, is given by
\begin{equation} \label{eq:1}
Y =  \frac{P_{hot}} {P_{cold}}
  = \frac{T_{hot}+T_{noise}}{T_{cold} + T_{noise}}
\end{equation}

\begin{equation} \label{eq:2}
T_{noise} = \frac{T_{hot} - YT_{cold}}{Y-1}
\end{equation}
where $T_{hot}$ and $T_{cold}$ are the temperatures of the reference sources and $T_{noise}$ is the equivalent input noise temperature of the device under test.

Prior to deployment in a Hot-Cold Testing Facility (HCTF), an initial hot-cold measurement was performed in-lab to verify the expected noise performance of the LNA. A single-polarization test circuit was constructed with a 130~$\Omega$ resistor connected at the differential input which corresponded approximately to $\Gamma_{opt}$ of the LNA. The resistor served as the noise source and was exposed to liquid nitrogen and ambient-temperature conditions. Power spectra were recorded for each state and analyzed offline using the Y-factor method. This measurement provided rapid in-lab validation of circuit functionality before undertaking more comprehensive noise measurements.

\begin{figure}[hbt!]
    \centering
    \includegraphics[width=0.82\linewidth]{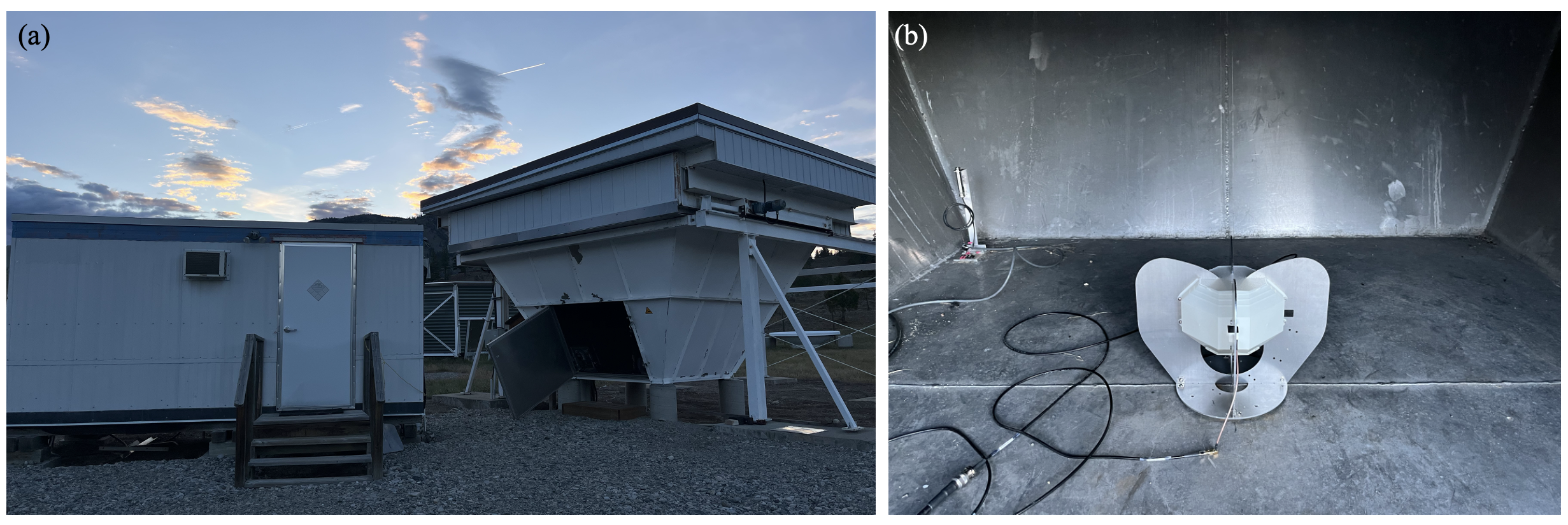}
    \caption{HCTF noise measurement setup. (a) Hot-Cold Test Facility consisting of a shielded test chamber with a retractable absorber roof and a control trailer housing the downstream receiver electronics and spectrum analyzers. (b) Vivaldi feed and differential LNA installed within the test chamber. The feed is supported by a custom injection-molded shroud\cite{lin2026} where the LNA is housed and coupled directly to the feed at the differential connection points.}
    \label{fig:hctf_setup}
\end{figure}

Detailed noise temperature measurements were subsequently performed using the HCTF located at the Dominion Radio Astrophysical Observatory (DRAO) \cite{hovey2018}. The HCTF is an automated measurement system consisting of a shielded test chamber with a retractable absorber roof that alternately provides hot and cold reference loads. When closed, the absorber acts as the hot source, while opening the roof exposes the system to the sky for cold-source measurements. For these tests, the dual-polarized LNA was mounted directly to the Vivaldi feed; The measurement set-up is shown in Figure~\ref{fig:hctf_setup}. Each polarization was measured while the other remained active, allowing any cross-coupling effects between channels to be captured. The outputs of the feed and LNA assembly were routed to the control trailer through 1 m sections of LMR-195 followed by 14 m of LMR-400 coaxial cable.

Within the trailer, the two chains (represented by each polarization) were connected to a custom RF enclosure containing downstream amplification and bias circuitry operating over a 300--1500 MHz frequency range. This stage provided approximately 5--30 dB of gain, increasing with frequency, and was designed to compensate for the gain slope of the LNA (Figure~\ref{fig:gain}) while supplying DC power to the front-end electronics. Following downstream amplification, each chain was connected to a spectrum analyzer for power measurements. Power spectra were integrated for three-minute intervals while alternating between hot (absorber-covered) and cold (sky-facing) states. Measurements were conducted over a two-hour period during nighttime conditions to minimize solar contributions to the cold reference temperature.

To determine the equivalent input noise temperature from the measured Y-factors, the cold-source temperature, $T_{cold}$, was modeled as a function of frequency rather than treated as a constant value. The Global Sky Model (GSM) was used to estimate the sky brightness temperature across the observing band and simulated beam patterns of the Vivaldi feed within the HCTF were applied to compute the beam-weighted sky temperature. This approach accounts for both the frequency dependence of the diffuse radio sky and the feed beam pattern, providing a more representative estimate of the effective cold-source temperature. The hot-source temperature, $T_{hot}$, was determined from an infrared temperature sensor integrated into the HCTF. Real-time measurements of the absorber surface temperature were recorded throughout the experiment and used directly in the Y-factor calculation. As the HCTF is designed and calibrated for operation above 600 MHz, measurements obtained below this frequency are subject to additional uncertainty and should be interpreted with caution.

\section{RESULTS}
\label{sec:results}

\subsection{Gain Characterization}

\begin{figure}
    \centering
    \includegraphics[width=0.8\linewidth]{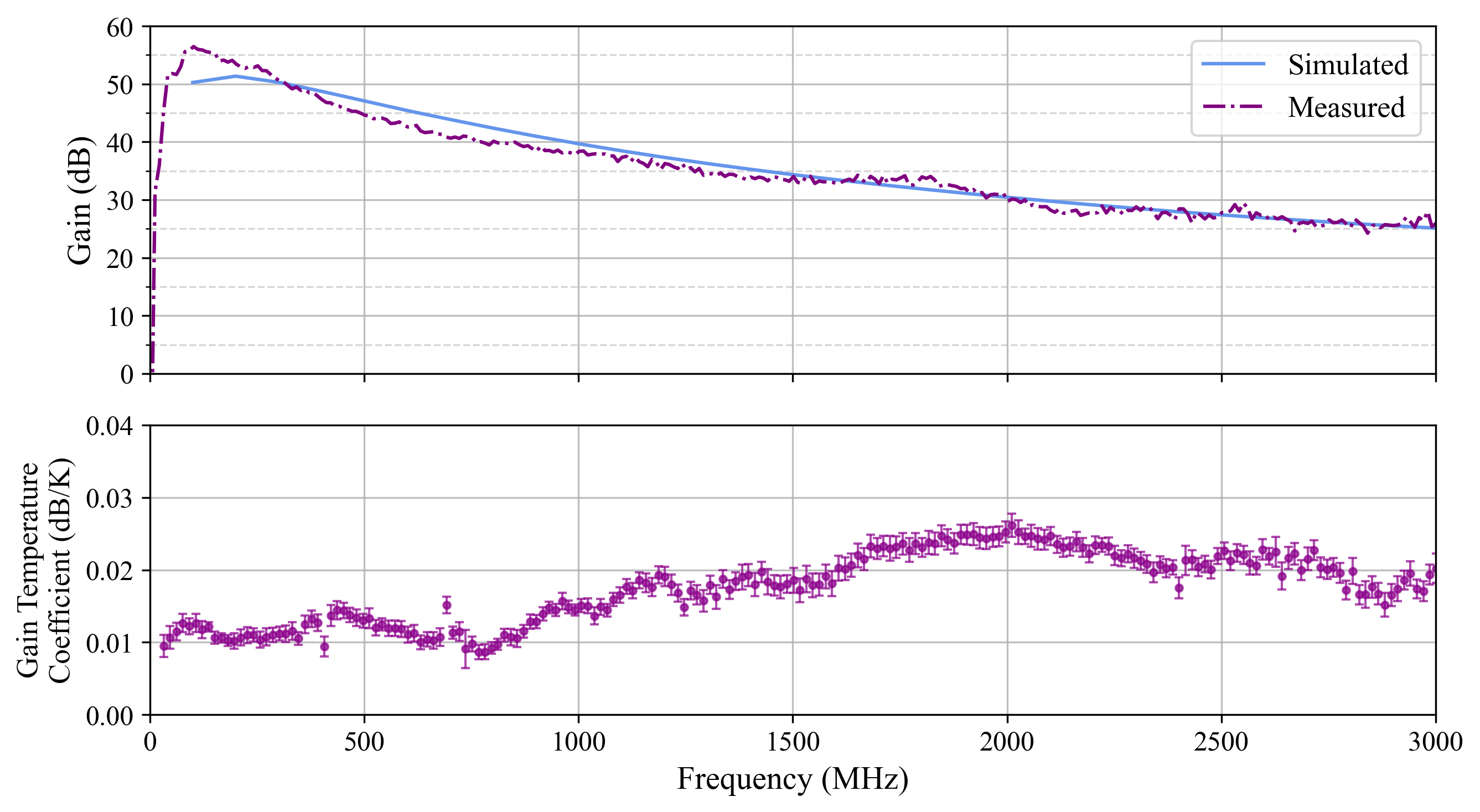}
    \caption{Measured gain ($S_{21}$) and GTC of the differential LNA prototype. (Top panel): Measured gain (purple, dashed), corrected for the insertion loss of the input balun, compared with the simulated response (blue). The amplifier provides more than 25 dB of gain over the measured 0.1--3 GHz frequency range. (Bottom panel): Measured GTC of the LNA, corrected for the independently characterized temperature-dependent insertion loss of the input balun. The GTC remains below 0.03 dB/K across the measured band and has a median value of 0.014 dB/K over the intended operating band of the Vivaldi feed (300--1500 MHz). Results are based on $S_{21}$ measurements spanning environmental temperatures of $-40^{\circ}\mathrm{C}$ to $60^{\circ}\mathrm{C}$.}
    \label{fig:gain}
\end{figure}

The measured gain of the differential LNA $-$ corrected for the insertion loss of the input balun $-$ is compared with simulated results in the upper panel of Figure~\ref{fig:gain}. Overall, the measured response is in good agreement with simulation over 0.1--3 GHz, with a gain $\geq25$ dB across this band. The LNA demonstrates an appreciable gain slope which is attributed to the frequency response of the QPL9547 and the second-stage amplifier. No equalization or band-limiting filters were incorporated into the current prototype; future revisions will integrate an equalizer to flatten the gain response and application-specific filtering to tailor the bandpass to the intended observing band. The lower panel of Figure~\ref{fig:gain} details the measured gain temperature coefficient (GTC) of the LNA after correcting for the independently characterized temperature-dependent insertion loss of the input balun. The GTC was determined by performing a least-squares fit to the measured gain as a function of temperature at each sampled frequency over the range of $-40^{\circ}\mathrm{C}$ to $60^{\circ}\mathrm{C}$, with the uncertainty estimated from the residuals of each fit. Across the 0.1--3 GHz band, the median gain temperature coefficient is 0.018 dB/K, decreasing to 0.014 dB/K over the intended operating band of the Vivaldi feed (300--1500 MHz).

\subsection{Noise Characterization}

\begin{figure}[hbt!]
    \centering
    \includegraphics[width=0.95\linewidth]{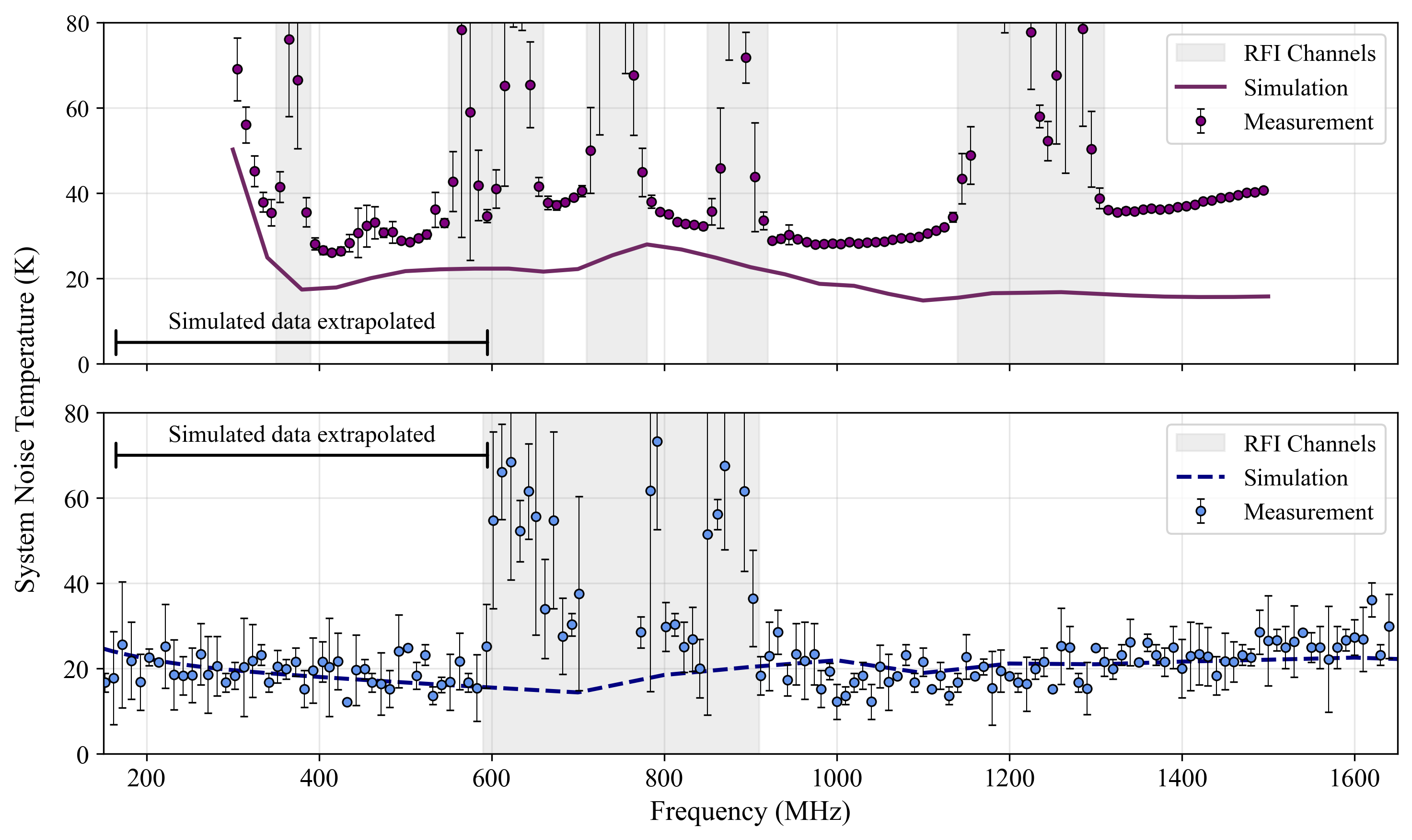}
    \caption{Measured and simulated noise temperature of the differential LNA prototype. (Top panel): Measured feed-coupled HCTF noise temperature (purple points) compared with the simulated response (purple line). The measured noise temperature reaches a minimum of approximately 25 K and remains below 40 K over the majority of the measured band. %Deviations from the simulated response are attributed to a combination of feed impedance mismatch, measurement uncertainty, and residual radio-frequency interference (RFI). 
    (Bottom panel): In-lab measured noise temperature of the LNA using a 130~$\Omega$ differential source impedance (blue points) compared with the simulated response (blue dashed line). The measurement validates the expected noise performance of the amplifier when presented with a source impedance close to its optimum differential noise match. For both plots, the frequency regions affected by persistent RFI are highlighted in grey and the simulated noise temperature is extrapolated below 600 MHz as manufacturer provided noise values are unavailable.}
    \label{fig:noise_result}
\end{figure}

The measured system noise temperatures obtained using the Vivaldi feed and a constant 130~$\Omega$ differential source are shown in Figure~\ref{fig:noise_result}, alongside the corresponding simulated responses. Both datasets were binned into 10 MHz frequency intervals with the error bars representing the standard deviation within each bin. 

\begin{figure}[hbt!]
    \centering
    \includegraphics[width=0.6\linewidth]{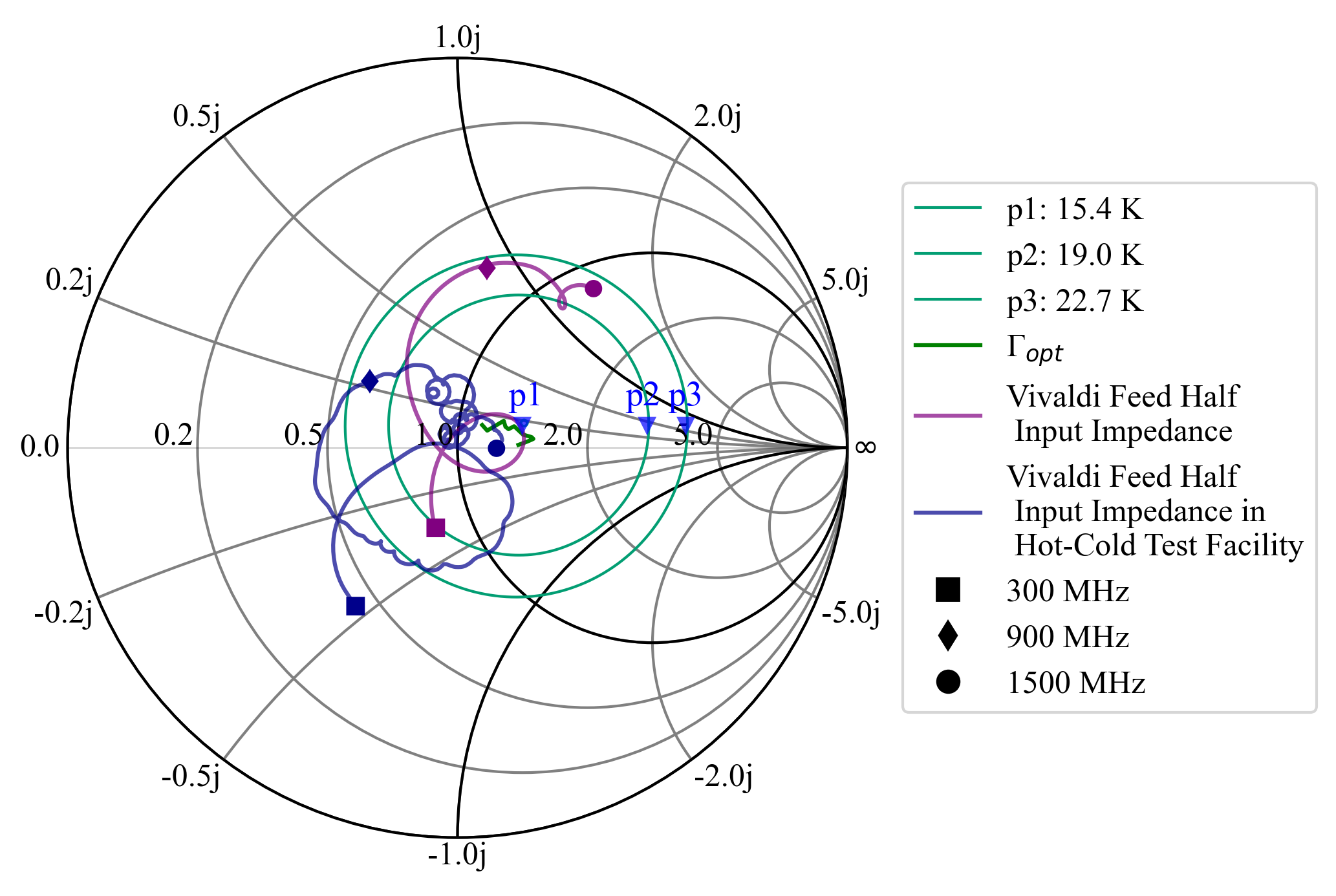}
    \caption{Smith chart illustrating the simulated differential input impedance of the Vivaldi feed in free space and in the HCTF compared to the optimal noise match of the QPL9547 LNA. (Purple) Simulated differential input impedance of Vivaldi feed halved, representing the reference match to one amplifier branch of the LNA. (Blue) Simulated differential input impedance of Vivaldi feed halved when placed in the HCTF chamber. (Turquoise) Noise circles of QPL9547 LNA at 900 MHz. (Green) $\Gamma_{opt}$ of QPL9547 across 300--1500 MHz band. (This data is presented on a Smith chart normalized to 50~$\Omega$).}
    \label{fig:noise_circ_HCTF}
\end{figure}

The upper panel of Figure~\ref{fig:noise_result} presents the feed-coupled HCTF measurements. The measured system noise temperature reaches a minimum of approximately 25 K and remains below 40 K over the majority of the operating band. The frequency regions exhibiting elevated and scattered system noise temperature are contaminated by persistent radio-frequency interference (RFI), primarily from GNSS and cellular bands which are highlighted in grey.  Overall, the measured response follows the trend predicted by simulation, although substantial deviations are observed at the upper end of the band. The simulated noise temperature falls below 20 K at higher frequencies due to the transformation of the feed input impedance when placed within the HCTF. Electromagnetic simulations of the complete feed-LNA assembly in the HCTF showed that this transformed impedance approaches the optimum noise match of the QPL9547 (see Figure~\ref{fig:noise_circ_HCTF}), resulting in a lower predicted system noise temperature. Although the measured data were corrected using a beam-weighted GSM to determine the effective cold-source temperature, several effects may contribute to the remaining discrepancy between measurement and simulation. These include residual reflections within the HCTF, uncertainties in the effective hot- and cold-source temperatures, and differences between the simulated and physical feed impedance. It is also of note that the HCTF is designed for operation above 600 MHz. While the measured response follows the simulated trend reasonably well below this frequency, the simulated noise temperature relies on extrapolation as manufacturer noise data for the QPL9547 are unavailable below 600 MHz.

The lower panel of Figure~\ref{fig:noise_result} presents the measured system noise temperature of the LNA when matched to a constant 130~$\Omega$ differential source. The measured response is in good agreement with simulation, validating the predicted noise performance of the amplifier near its optimum differential noise match. The measurement was conducted in a lab environment in downtown Toronto where appreciable RFI is present particularly from local cellular bands which explains the increase in system noise temperature close to 700 MHz.

\section{Discussion \& CONCLUSION}\label{sec:discuss}

We have developed a low-cost, ultra-wideband, dual-polarized differential LNA designed for differential feeds used in radio astronomy. Fabricated entirely from commercially available surface-mount components, the design provides a practical and scalable front-end solution for next-generation large-N radio telescopes. By eliminating the need for an input balun, the architecture simplifies the receiver front-end while maintaining the low-noise performance sought after by modern radio astronomy instruments. Although demonstrated using a 300--1500 MHz Vivaldi feed, the LNA is intended as a general-purpose resource. The circuitry supports substantially wider bandwidths than demonstrated in this work, while the use of commercially available components and a modular architecture enables adaptation to a wide range of differential feed interfaces and radio astronomy applications. The design is currently being evaluated across several projects including a prototype ionospheric monitor for the Hydrogen Intensity and Real-time Analysis eXperiment (HIRAX), CHORD outrigger receiver chain development for fast radio burst localization, and a receiver for the 46-m dish at the Algonquin Radio Observatory supporting pulsar scintillation studies. A closely related variation of the architecture has also been implemented on the active antenna for the Canadian-Chilean Array for Radio Transient Studies (CHARTS)~\cite{lau2025, cassanelli2025}. Ongoing work is focused on integrating an on-board equalizer to flatten the broadband gain response and exploring options for application-specific filtering tailored to individual observing bands. In parallel, three prototype LNAs have been deployed on the CHORD Pathfinder, where on-sky measurements are underway to characterize their noise performance. 

This work demonstrates that high-performance differential front-end electronics can be realized using low-cost commercial components, providing a practical solution for future radio astronomy instrumentation.

%%%%%%%%%%%%%%%%%%%%%%%%%%%%%%%%%
\acknowledgments      
The Dunlap Institute is funded through an endowment established by the David Dunlap family and the University of Toronto.
The authors acknowledge the support of the Natural Sciences and Engineering Research Council of Canada (NSERC). We also thank the staff at the Dominion Radio Astrophysical Observatory for providing access to the Hot-Cold Test Facility, which enabled the noise measurements presented in this work. We also acknowledge the CHORD Collaboration for providing the infrastructure and technical resources supporting the ongoing testing and characterization of the current LNA design.

% References
\bibliography{report} % bibliography data in report.bib
\bibliographystyle{spiebib} % makes bibtex use spiebib.bst

\newpage
%%%%%%
\appendix    %>>>> this command starts appendixes

\section{Input-referenced Noise in Ideal Differential Systems}\label{sec:noise_derivation}

Here it is shown that the input-referenced noise of parallel LNAs is that of a single-ended LNA assuming the parallel LNAs share the same noise, gains and operate $180$ degrees out of phase. For clarity this derivation also assumes an ideal differential input signal and electronic system, neglecting common-mode response effects. 

\subsection{System Layout}
\begin{figure}[hbt!]
    \centering
    \includegraphics[width=0.65\linewidth]{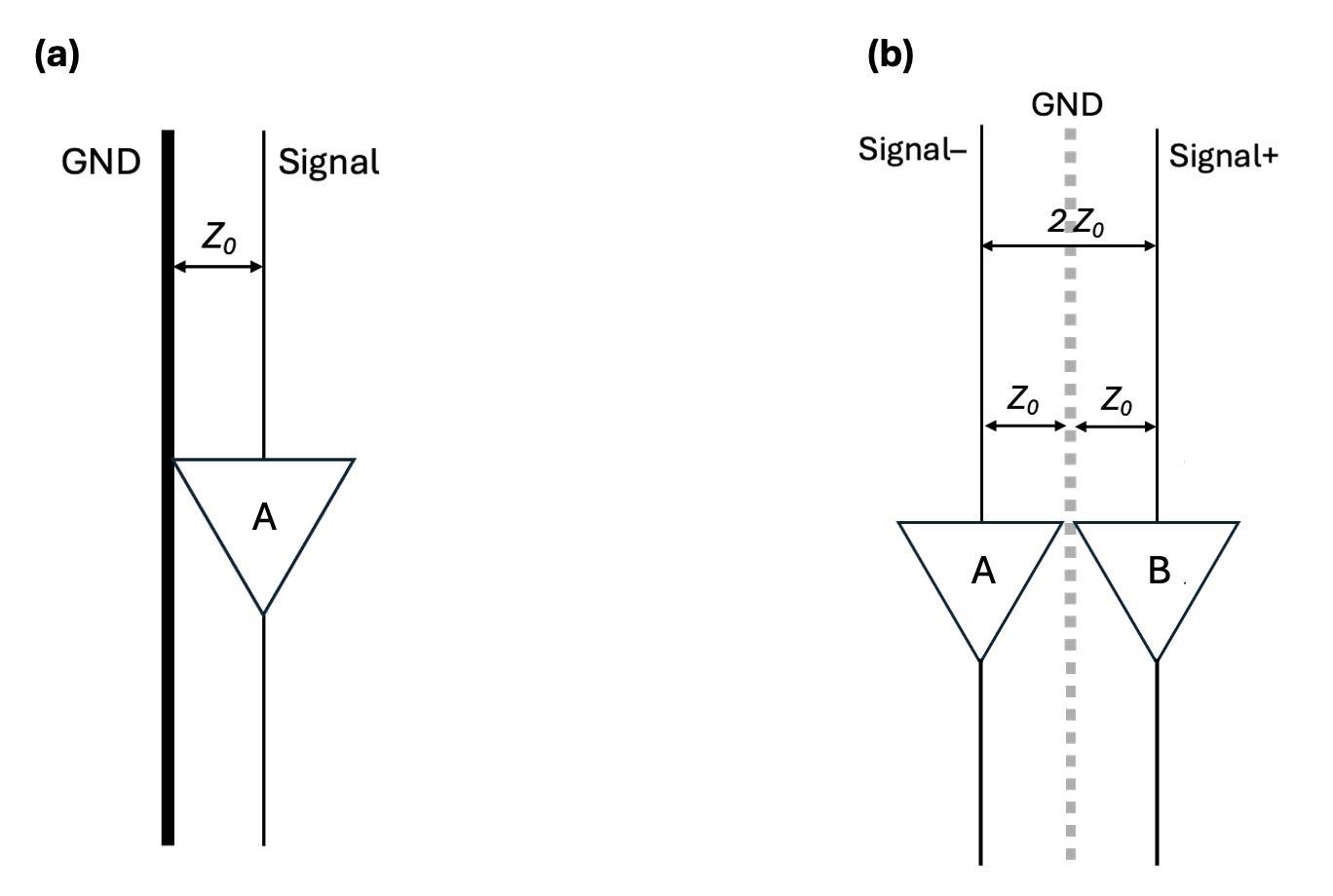}
    \caption{System layout of (a) single-ended LNA and (b) parallel LNAs. The single-ended LNA is matched to an antenna supplying $Z_0$ source impedance whereas the parallel LNAs are being supplied a source impedance of $2Z_0$ from a differential feed with each polarity presenting $Z_0$ to ground.}
    \label{fig:LNA_system}
\end{figure}

This derivation compares a single-ended LNA topology to a differential one. The differential system is composed of two LNAs placed in parallel, each on independent traces that share a common ground. The traces can then be later combined with a balun or directly fed to a differential readout. Figure~\ref{fig:LNA_system} details these systems and provides a simplified description of the input impedance for each.

\subsection{Single-ended and Differential Noise}

We start with the idea that the power, $P$ and effective temperature, $T$, of a single-ended system can be expressed in terms of the instantaneous voltage, $V$, drawn randomly from a Gaussian distribution divided by the system’s impedance, $Z_0$.

\begin{equation}\label{eq:temp_LNA}
    T \propto P \propto \frac{Var(V)}{Z_0}
\end{equation}

The variance in the voltage can be written out as the average of the voltage squared subtracted by the average voltage squared. Given this derivation is concerned with the propagation of AC signals, the average voltage is taken as 0 such that, 

\begin{equation}
\begin{split}
    Var(V) &= \langle |V|^2\rangle - \cancel{|\langle V \rangle|^2}\\
    & = \langle |V|^2\rangle
\end{split}
\end{equation}

For a standard single-ended system, there is a single RF transmission line that carries the absolute signal throughout the chain. The total power of the system, $P_{sing,tot}$, can be (simply) broken down into two components $-$ the received power from the sky, $P_{s}$ (with temperature $T_{s}$), and the power from the electronics along this chain, in this case we will isolate a single-ended LNA with power $P_{n}$.

\begin{equation}\label{eq:power_vars_single}
    P_{sing,tot} = P_{s} + P_{n}
\end{equation}

The sky voltage can then be expressed such that,

\begin{equation}
    \langle |V_{s}|^2 \rangle = T_{s}Z_0
\end{equation}

In terms of the parallel LNAs, in this differential system, the RF signal is contained by two polarities that propagate along two independent conductors whilst sharing a common ground. Given these two polarities (or two “branches”), the impedance of the system is two times the impedance of the single-ended system ($2Z_0$). Assuming the differential system observes the same sky temperature as a single-ended system, the sky voltage is defined by,

\begin{equation} \label{eq:imped1}
\begin{split}
    \langle |V_{diff,s}|^2 \rangle &= T_{s}(2Z_0) \\
    & = 2\langle |V_{s}|^2 \rangle
\end{split}
\end{equation}

The differential sky voltage contributes to two polarities 180 degrees out of phase such that $V_{diff,s} = 2V_{pol,s}$. Therefore we can expand the differential sky signal and write it in terms of a single polarity, 

\begin{equation}\label{eq:imped2}
\begin{split}
    \langle |V_{diff,s}|^2 \rangle &= \langle |2V_{pol,s}|^2 \rangle \\
    &= 4\langle |V_{pol,s}|^2 \rangle
\end{split}
\end{equation}

Equating Equations \ref{eq:imped1} and \ref{eq:imped2} gives,

\begin{equation}\label{eq:pol_to_sky}
    \langle |V_{pol,s}|^2 \rangle = \frac{\langle |V_{s}|^2 \rangle}{2}
\end{equation}

where $V_{pol,s}$ is the sky voltage seen by one polarity.

Each LNA amplifying its associated polarity will contribute to the noise as per Equation~\ref{eq:temp_LNA}. Note that when modeling each polarity, the impedance is half of the total system impedance ($2Z_0$) therefore it is taken as $Z_0$ based on Figure~\ref{fig:LNA_system}.

\begin{equation}
    \langle |V_{pol,n}|^2 \rangle = T_{n}Z_0
\end{equation}

Combining both the sky and noise contributions for each polarity, the total voltage ($V_{pol,tot}$) is, 

\begin{equation}
    \langle |V_{pol,tot}|^2 \rangle = \langle |V_{pol,s}  + V_{pol,n} |^2 \rangle
\end{equation}

Expanding this out leaves the following result as the cross term cancels due to the sky signal and noise of the LNA being uncorrelated. 
\begin{equation}
\begin{split}
    \langle |V_{pol,tot}|^2 \rangle 
    &= \langle |V_{pol,s}|^2\rangle  +\langle| V_{pol,n} |^2 \rangle +  \cancel{2\langle| V_{pol,s}V_{pol,n} | \rangle}\\
    &= \langle |V_{pol,s}|^2\rangle  +\langle| V_{pol,n} |^2 \rangle
\end{split}
\end{equation}

Now, to account for the \textbf{total} voltage of the differential system, the two polarities (1 and 2) are added, 

\begin{equation}
    \begin{split}
        \langle |V_{diff,tot}|^2 \rangle 
        &= \langle |V_{pol1,tot} + V_{pol2,tot}|^2 \rangle
    \end{split}
\end{equation}

\begin{equation}
    \begin{split}
        \langle |V_{diff,tot}|^2 \rangle 
        &= \langle |V_{pol1,s} + V_{pol1,n} + V_{pol2,s} + V_{pol2,n}|^2 \rangle
    \end{split}
\end{equation}

Expanding and canceling terms that go to zero, 

\begin{equation}
    \begin{split}
        \langle |V_{diff,tot}|^2 \rangle 
        &= \langle |V_{pol1,s}|^2 \rangle + \langle|V_{pol1,n}|^2 \rangle + \langle |V_{pol2,s}|^2\rangle + \langle|V_{pol2,n}|^2 \rangle + 2\langle |V_{pol1,s}V_{pol2,s}| \rangle  \\
        & 
    \end{split}
\end{equation}

Note that $V_{pol1,s}$ and $V_{pol2,s}$ are equal to the generic term $V_{pol,s}$ and add coherently. All other cross terms cancel.

\begin{equation}
    \begin{split}
        \langle |V_{diff,tot}|^2 \rangle 
        = 4\langle |V_{pol,s}|^2 \rangle + \langle|V_{pol1,n}|^2 \rangle  + \langle|V_{pol2,n}|^2\rangle   \\
    \end{split}
\end{equation}

Plugging in Equation~\ref{eq:pol_to_sky} for the sky term gives, 

\begin{equation}
        \langle |V_{diff,tot}|^2 \rangle 
         = 2\langle |V_{s}|^2 \rangle  + \langle|V_{pol1,n}|^2 \rangle  + \langle|V_{pol2,n}|^2\rangle  
\end{equation}

Now converting the total differential voltage into total differential power using Equation~\ref{eq:temp_LNA} and setting the impedance to $2Z_0$ for the whole system gives the expression: 

\begin{equation}
    P_{diff,tot} = \frac{2\langle |V_{s}|^2 \rangle  + \langle|V_{pol1,n}|^2 \rangle  + \langle|V_{pol2,n}|^2\rangle}{2Z_0}
\end{equation}

Expanding the relation out and converting all terms into power space, 

\begin{equation}
\begin{split}
    P_{diff,tot} 
    &= \frac{\langle |V_{s}|^2 \rangle}{Z_0}  + \frac{\langle|V_{pol1,n}|^2 \rangle}{2Z_0}  + \frac{\langle|V_{pol2,n}|^2\rangle}{2Z_0}\\
    &=P_{s}  + \frac{P_{pol1,n}}{2}  + \frac{P_{pol2,n}}{2}
\end{split}
\end{equation}

Assuming the power from each LNA along either polarity is equal, the result is,
\begin{equation}
    P_{diff,tot} = P_{s} + P_{n}
\end{equation}

Meaning that the total power (or effective temperature) of a differential system with parallel LNAs is the same as a single-ended version (Equation~\ref{eq:power_vars_single}),

\begin{equation}
    \boxed{P_{sing,tot} = P_{diff,tot}}
\end{equation}

With this relation, the voltages of both the differential system and single-ended system can be related such that 
\begin{equation}
\begin{split}
    & \frac{\langle|V_{diff,tot}|^2\rangle}{2Z_0} = \frac{\langle|V_{sing,tot}|^2\rangle}{Z_0}\\
    & \langle|V_{diff,tot}|^2\rangle = 2\langle|V_{sing,tot}|^2\rangle
\end{split}
\end{equation}

Or more clearly, 
\begin{equation}
    \boxed{V_{diff,tot} = \sqrt2 V_{sing,tot}}
\end{equation}

Which means that each polarity or "branch" of the differential signal can be represented as $1/\sqrt2$ of the single-ended signal in voltage space.

\begin{equation}
    \frac{V_{diff,tot}}{2} = \frac{V_{single,tot}}{\sqrt2}
\end{equation}

%%%%%%%%%%%%%%%%%%%%%%%%%%%%%%%%%%%%%%%%%%%%%
\section{Signal to Noise Ratio of Differential Systems}\label{sec:snr_derivation}

This appendix builds upon the derivation presented in Appendix~\ref{sec:noise_derivation} by comparing the SNR of differential and single-ended LNA topologies (as shown in Figure~\ref{fig:LNA_system}). The analysis accounts for gain and phase mismatches arising from component tolerances and manufacturing variations, providing a more realistic assessment of differential amplifier performance. This derivation assumes an ideal differential input signal and electronic system, neglecting common-mode response effects. 

%%%%%%%
\subsection{Single-ended System: Signal to Noise Ratio} \label{sec:single_derivation}
The sky voltage, $V_{s}$, is a complex random variable whose probability distribution is Gaussian with zero mean and variance, $\sigma_{s}^2$, equal to the sky temperature, $T_{s}$. For simplicity the mean phase is set to zero, meaning $V_{s}$ can be treated as real-valued fluctuations.

\begin{equation}\label{vsky}
    \langle|V_{s}|^2\rangle = \sigma^2_{s} = T_{s}
\end{equation}

The LNA, denoted A, has a complex linear gain,
\begin{equation}\label{gains_sing}
    G_{A}=Ae^{i\phi}
\end{equation}

where $A \geq 0$ is a magnitude and $\phi$ is the phase. 

The LNA contributes its own additive noise which is described as,
\begin{equation}\label{sing_noise}
    \langle|V_{n}|^2\rangle = \sigma_n^2 = T_n
\end{equation}
%%%%%%

\subsubsection{Signal contribution}

Isolating the resulting signal from the system ($V_{sig}$), the output from the LNA is,

\begin{equation}
    V_{sig} = Ae^{i\phi}V_{s} 
\end{equation}

The signal power is proportional to,
\begin{equation}
\begin{split}
    P &\propto Var(V_{sig})\\
     &\propto \langle|V_{sig}|^2\rangle - \cancel{|\langle V_{sig}\rangle|^2}
\end{split}
\end{equation}

Expanding this provides a final expression for the signal power,

\begin{equation}\label{sing_finalsignal}
    \boxed{\langle|V_{sig}|^2\rangle = A^2 \langle|V_{s}|^2\rangle}
\end{equation}

%%%%%
\subsubsection{Noise contribution}

The noise contribution of the LNA is modeled using Equation \ref{sing_noise}. The noise output from the single-ended LNA is,

\begin{equation}
    V_{noise} = AV_{n}
\end{equation}

Which can be expressed as,

\begin{equation}\label{sing_finalnoise}
    \boxed{\langle |V_{noise}|^2\rangle = A^2\langle |V_{n}|^2 \rangle}
\end{equation}

or, 

\begin{equation}
    \langle |V_{noise}|^2\rangle = A^2T_A
\end{equation}

%%%%%
\subsubsection{Signal to noise ratio}
Although trivial, the resulting SNR for this system is written for completeness. This is done by substituting in Equations \ref{sing_finalsignal} and \ref{sing_finalnoise}.
\begin{equation}
    SNR_0 = \frac{\langle|V_{sig}|^2\rangle}{\langle |V_{noise}|^2\rangle}
\end{equation}

\begin{equation}
    SNR_0 = \frac{A^2 \langle|V_{s}|^2\rangle}{A^2\langle |V_{n}|^2 \rangle}
\end{equation}

\begin{equation}\label{single_SNR}
    \boxed{SNR_0 = \frac{\langle|V_{s}|^2\rangle}{\langle |V_{n}|^2 \rangle}}
\end{equation}

%%%%%
\subsection{Differential System: Signal to Noise Ratio}

In this section, a SNR calculation is repeated for a differential topology with layout as per Figure~\ref{fig:LNA_system}(b) In this case, each polarity of the differential signal is amplified independently against a common ground by a single-ended LNA. The sky voltage follows the same form as the single-ended case presented in Appendix \ref{sec:single_derivation} however, given that each differential leg equates to $\frac{1}{\sqrt2}$ of the absolute signal in voltage space (derived in Appendix \ref{sec:noise_derivation}), then, in this case, each LNA or polarity will observe $\frac{V_{s}}{\sqrt2}$.
\\

Each LNA, denoted A and B, has a complex linear gain,
\begin{equation}\label{gains}
    G_{A}=Ae^{i\phi_{A}}, G_{B} = Be^{i\phi_{B}} 
\end{equation}

where $A,B \geq 0$ are magnitudes and $\phi_{A,B}$ are phases. 

Each LNA contributes its own additive noise $n_{A}$ and $n_{B}$ $-$ complex random variables with zero mean and variance described as,
\begin{equation}\label{LNAnoise}
\begin{split}
    \langle|V_{n_{A}}|^2\rangle = \sigma_{A}^2 = T_A &\\ 
    \langle|V_{n_{B}}|^2\rangle = \sigma^2_{B} = T_{B}
\end{split}
\end{equation}

Note that LNA noise contributions are uncorrelated as we are accounting for the internal noise figures only. 

%%%%
\subsubsection{Signal contribution}
%%%%
Isolating the resulting signal from the system, the output from LNA A is,
\begin{equation}
    V_{A,sig} = Ae^{i\phi_A}\frac{1}{\sqrt2}V_{s} 
\end{equation}

and the signal output from LNA B,
\begin{equation}
    V_{B,sig} = Be^{i\phi_B}\frac{1}{\sqrt2}V_{s}
\end{equation}

The outputs from each LNA can be modeled as Gaussian random variables with zero mean and variances given by
\begin{equation}
\begin{split}
   V_{A,sig} \approx \mathcal{G}(0, A^2\sigma_{s}^2 ) & \\
   V_{B,sig} \approx \mathcal{G}(0, B^2\sigma_{s}^2)
\end{split} 
\end{equation}

$V_{A,sig}$ and $V_{B,sig}$ are individually Gaussian and statistically correlated due to the shared sky contribution. They make up linear combinations of the Gaussian random variable $V_{s}$, meaning they are jointly Gaussian. Therefore, their linear combination $V_{sig} = V_{A,sig} + V_{B,sig}$ is also a Gaussian random variable.\\

Combining the outputs of both LNAs provides an expression for the total voltage of the signal, $V_{sig}$, 

\begin{equation}
\begin{split}
   V_{sig} &= V_{A,sig} + V_{B,sig}\\
   &=\frac{1}{\sqrt2}V_{s}(Ae^{i\phi_A} + Be^{i\phi_B}) 
\end{split} 
\end{equation}

The signal power is proportional to,
\begin{equation}
\begin{split}
    P &\propto Var(V_{sig})\\
     &\propto \langle|V_{sig}|^2\rangle - \cancel{|\langle V_{sig}\rangle|^2} \\
\end{split}
\end{equation}

Where, 
\begin{equation}
\begin{split}
    \langle|V_{sig}|^2\rangle & = \langle|\frac{1}{\sqrt2}V_{s}(Ae^{i\phi_A} + Be^{i\phi_B}) |^2\rangle\\
     & =|Ae^{i\phi_A} + Be^{i\phi_B}|^2\frac{1}{2}\langle|V_{s}|^2\rangle 
\end{split}
\end{equation}

Expanding out the complex gain term,
\begin{equation}
\begin{split}
    |Ae^{i\phi_A} + Be^{i\phi_B}|^2 &= |A|^2 + |B|^2 + 2Re(Ae^{i\phi_{A}}(Be^{i\phi_{B}})^*)\\
    &= A^2 + B^2 + (2AB)Re(e^{i(\phi_{A}-\phi_{B})})\\
    &= A^2 + B^2 + 2ABcos(\phi_{A}-\phi_{B})
\end{split}
\end{equation}

Plugging this result back in, the signal power is now proportional to,

\begin{equation}
    \langle|V_{sig}|^2\rangle = \bigl[A^2 + B^2 + 2ABcos(\phi_{A}-\phi_{B})\bigr] \frac{1}{2}\langle|V_{s}|^2\rangle 
\end{equation}

To simplify the expression, set 
\begin{equation}
    \begin{split}
        &\gamma = \phi_{A} - \phi_{B} \\
        &\alpha = \frac{B}{A} \\
        %&\beta = \frac{T_B}{T_A}
    \end{split}
\end{equation}

The relation then becomes 
\begin{equation}\label{diff_signal}
    \boxed{\langle|V_{sig}|^2\rangle = (1 + \alpha^2 + 2\alpha  cos(\gamma))\frac{1}{2}A^2\langle|V_{s}|^2\rangle}
\end{equation}

Using Equation \ref{vsky}, this result can also be written in terms of the sky temperature: 

\begin{equation}
    \langle|V_{sig}|^2\rangle = (1 + \alpha^2 + 2\alpha  cos(\gamma))\frac{1}{2}A^2T_{s}
\end{equation}

%%%%
\subsubsection{Noise contribution}
%%%%
The noise contribution of each LNA is modeled using Equation \ref{LNAnoise}. Isolating the individual systems, the noise output from LNA A is,

\begin{equation}
    V_{A,noise} = AV_{n_A}
\end{equation}

and the output from LNA B is, 

\begin{equation}
    V_{B,noise} = BV_{n_B}
\end{equation}

Given that the noise contributions of LNAs A and B are uncorrelated, the cross terms cancel and the resulting noise is represented by,

\begin{equation}
\begin{split}
    \langle|V_{noise}|^2\rangle & = \langle|V_{A,noise} + V_{B,noise}|^2\rangle \\ 
    &= A^2\langle|V_{n_A}|^2\rangle + B^2\langle|V_{n_B}|^2\rangle \\
    & = A^2T_A + B^2T_B
\end{split}
\end{equation}

Setting $\beta = \frac{T_B}{T_A}$, and using expressions for $\alpha$ and $\gamma$, gives a simplified expression:

\begin{equation}\label{diff_noise}
\boxed{
\begin{split}
    \langle|V_{noise}|^2\rangle &= (1+\alpha^2\beta)A^2T_A\\
    &=(1+\alpha^2\beta)A^2\langle|V_{n_A}|^2\rangle
\end{split}
    }
\end{equation}

%%%%
\subsubsection{Signal to noise ratio}
%%%%

Given the derivation of the differential signal (Equation \ref{diff_signal}) and noise (Equation \ref{diff_noise}), one can express the SNR of the parallel LNA system as,

\begin{equation}\label{init_SNR}
\begin{split}
    SNR^2 &= \frac{\langle|V_{sig}|^2\rangle}{\langle|V_{noise}|^2\rangle}\\
    &=\frac{(1 + \alpha^2 + 2\alpha  cos(\gamma))\langle|V_{s}|^2\rangle}{2(1+\alpha^2\beta)\langle|V_{n_A}|^2\rangle}
\end{split}
\end{equation}

Taking the SNR from the single-ended case (Equation \ref{single_SNR}) and substituting this in (where $V_n = V_{n_A}$) gives, 

\begin{equation}\label{init_SNR}
\begin{split}
    \boxed{SNR^2 =SNR_0^2\frac{(1 + \alpha^2 + 2\alpha  cos(\gamma))}{2(1+\alpha^2\beta)}}
\end{split}
\end{equation}

This equation describes the impact of LNA mismatch on the given systems SNR. If all variables such as gains, noise and phase agree between the parallel LNAs (i.e. $\alpha = 1, \beta = 1, \phi = 0$), then $\text{SNR} = \text{SNR}_0$ as expected. If there is an associated mismatch between these variables, Figure~\ref{fig:snr_deg} describes the expected behavior of the SNR.

\end{document}